# Fluxiod jumps coupled high critical current density of nano-$Co_3O_4$ doped $MgB_2$

*V.P.S. Awana[1,2,$], M. Isobe[2], K.P. Singh[1,3], Md. Shahabuddin[3], H. Kishan[1], and E. Takayama-Muromachi[2]*

*[1]National Physical Laboratory, K.S. Krishnan Marg, New Delhi-110012, India*

*[2]Superconducting Materials Center, NIMS, 1-1 Namiki, Tsukuba, Ibaraki, 305-0044, Japan*

*[3] Physics Department, Jamia Millia Islamia University, New Delhi 110025, India*

Polycrystalline $MgB_2$ samples with 0, 2, 4 and 6% added nano-$Co_3O_4$ being synthesized by vacuum ($10^{-5}$ Torr) annealing at 750 $^0$C for three hours each are found to be nearly single phase with presence of only a small quantity of Mg/MgO in pristine sample and in addition the $Co_2O_3$ in doped compounds. All the samples exhibited clear and sharp diamagnetic transitions at around 38 K, in Zero-field-cooled (*ZFC*) magnetic susceptibility measurements with sizeable signal. The Field cooled (*FC*) measurements though having sharp transitions, but showed a very small signal, indicating high level of pinning centers in these samples. Further some of the doped samples exhibited Paramagnetic-Meissner-Effect (*PME*) in applied field of 5 Oe. The critical current density ($J_c$), being estimated by invoking Bean model for the pristine compound increase by nearly an order of magnitude for 2% and 4% nano-$Co_3O_4$ doping and later the same decreases sharply for 6% sample at nearly all studied temperatures and applied fields. Further the increased $J_c$ (~ $10^8$ A/cm$^2$) is coupled with fluxiod jumps ($T \leq 20$ K and H $\leq$ 1 T). Fluxiod Jumps are not seen in relatively low $J_c$ pristine or 6% sample. This means the fluxiod-jumps are intrinsic only to the high $J_c$ samples.

** Corresponding author: e.mail- awana@mail.nplindia.ernet.in, and AWANA.veer@nims.go.jp*

## INTRODUCTION

Recent observations of the relatively high critical density of $MgB_2$ had attracted a lot of attention [1,2]. Fundamentally besides the superconducting critical temperature ($T_c$) of 40 K, the relatively higher coherence length (~ 50 *nm*) in comparison to high temperature cuprate superconductors (HTSc) made $MgB_2$ very susceptible for doped nano-particles to act as pinning centers. For any impurity to act as pinning centers in a superconductor its size and distribution in the host are very important. The superconducting coherence length of $MgB_2$ of up to 50 *nm* permits most of the doped nano-particles to pin the vortices and hence an increased $J_c$, by avoiding vortex resistive flow in the material.

By now a number of articles had appeared on the critical current density of nano particle doped $MgB_2$ for example nano-$SiO_2$ [3], nano-Fe [4], nano-diamond [5], nano-SiC [6], nano-$TiO_2$ [7], nano-$Al_2O_3$ [8] and others. Not only the doped nano-particles but various routes of synthesis had also resulted in improved critical current density of this compound [9,10]. Very recently we reported the enhanced critical current density (> $10^7$ A/cm$^2$) coupled with fluxiod-jumps for a vacuum-annealed and Mg-turnings added $MgB_2$ [11]. Interestingly this compound consisted reasonable amount of Mg/MgO in it [11]. Not only the amount of added Mg turnings but the heating rate also controls the amount of Mg/MgO in the end product. In the present study instead of using Mg/MgO as pinning centers in $MgB_2$, we added nano-$Co_3O_4$ in the pristine $MgB_2$ and found an enhanced $J_c$. Though in the present study the Mg turnings are not added, still comparatively [11] small but some Mg/MgO is seen in XRD (x-ray diffraction) of the all samples of present study also. Further we found that the increased $J_c$ (~ $10^8$ A/cm$^2$) of the nano-$Co_3O_4$ particles doped $MgB_2$ is coupled with fluxiod jumps ($T \leq 20$ K and H $\leq$ 1 T) similar to that as for Mg/MgO included samples [11]. In fact the $J_c$ (~ $10^8$ A/cm$^2$) at $T \leq 20$ K and H $\leq$ 1 T can be considered as nearly invariant similar to that as found more precisely in ref.11. The importance of fluxiod jumps in relation to high and nearly invariant $J_c$ at lower temperatures (< 10 K) and applied fields (< 1T) will be discussed in detail in next sections.

We also compare our present results of nano-$Co_3O_4$ doped $MgB_2$ with various other reports [3-8] on different nano-particle doped $MgB_2$. We found that in most of the nano-particle doped $MgB_2$, the high $J_c$ is coupled with low field and low temperature fluxiod jumps. The phenomenon of fluxiod-jumps seems to be intrinsic to the high pinning density nano-particles doped $MgB_2$ compounds which invariably possess high $J_c$. We report in this article one more such example of nano-$Co_3O_4$ doped $MgB_2$.

## EXPERIMENTAL DETAILS

$MgB_2$ samples with 0, 2, 4 and 6% added nano-$Co_3O_4$ (particle size 20*nm*) samples were synthesized by encapsulation of well mixed and pelletized high quality (above 3 N purity) Mg, B and nano-$Co_3O_4$ powders in a soft Fe-tube and its subsequent heating to 750 $^0$C for two and half hours in an evacuated ($10^{-5}$ Torr) quartz tube and quenching to liquid-nitrogen temperature. The nano- $Co_3O_4$ is used precisely due to the fact, so that it could pin the vortices, which are typically double to the coherence length (~ 50 *nm*) of the $MgB_2$. In our synthesis the Fe-encapsulation is mainly for safety reasons and to avoid direct contact of the samples with quartz. The resultant samples are bulk polycrystalline quite porous compounds. The detailed method of synthesis is described elsewhere, please see ref. 12. The x-ray diffraction pattern of the compound was recorded with a diffractometer using $CuK_\alpha$ radiation. Magnetization measurements were carried out with a Quantum-Design SQUID magnetometer *(MPMS-XL).* All the samples for magnetization measurements were in more or less same size rod like shapes. The mass of the samples was kept around 10-15 mg. each, to keep the output magnetization signal of the machine below its dynamical limit of ± 5 emu. This is important to mention because some of our samples showed strong intrinsic fluxiod jumps in magnetization.

## RESULTS AND DISCUSSION

Resistiviy ($\rho$) versus temperature ($T$) plots for 0, 2, 4 and 6% nano-$Co_3O_4$ doped $MgB_2$ system are shown in Figure 1. It is seen that all the samples exhibit metallic resistivity down to say 40 K and a sharp a superconducting transition to resistiviy ($\rho$) = 0 state at 37-38 K. This does mean that the superconducting transition temperatue ($T_c$) is not effected much with addition of $Co_3O_4$ in the polycrystalline $MgB_2$. The $\rho$ ($T$) has a near constant metallic slope down to say 100 K, and later it seems to follow power law with less positive slope. The typical $\rho$ ($T$) of $MgB_2$ had earlier been explained by some us in detail [13]. The typical $\rho$ ($T$) behavior is similar to earlier reports on various $MgB_2$ compounds [14, 15]. Normal state resistivity increases with increase in $Co_3O_4$ content in the samples, indicating an increased disorder.

Figure 2 depicts the DC magnetic susceptibility ($\chi$) versus temperature ($T$) plots in both Zero-field-cooled (*ZFC*) and Field-cooled (*FC*) situations for 0, 2, 4 and 6% nano-$Co_3O_4$ added $MgB_2$ system. The applied field is 5 Oe. The *ZFC* branch of the magnetization undergoes a sharp diamagnetic transition below 39 K (*$T_c$)* for all the samples, irrespective of the $Co_3O_4$ content.

This shows that Co is not substituted at either Mg or B site in the $MgB_2$. In fact substitution of magnetic Co at Mg site in $Mg_{1-x}Co_xB_2$ is supposed to decrease $T_c$ [16]. Apparently that is not the situation in present case. As far as the *FC* branch of the magnetization is concerned, the same shows paramagnetic-Meissner (*PME*) like transition below $T_c$ for all the samples. It is known earlier that in highly pinned $MgB_2$ samples, the *FC* transition of magnetization is very weak. The appearance of *PME* is most probably the indication of stacked intrinsic SIS (Superconductor-Insulator-Superconductor) or SNS (Superconductor-Normal-Superconductor) junctions at micron level [17, 18]. The pi-junctions of SIS/SNS in favorable conditions can give rise to *PME*. Observation of *PME* indicates the presence of nano-metric normal/insulating impurities in our compound. In fact, we found some evidence for this in our x-ray diffraction studies, to be discussed in the last. As far as the volume fraction of the superconductivity is concerned, it is difficult to ascertain the same in the presence of possible high pinning and the *PME* phenomenon in the present samples. Worth noting is the fact that in various earlier reports [7-11], a similar behavior of *FC* magnetization in terms of either very weak diamagnetism or *PME* phenomenon is seen in highly pinned nano-particles doped $MgB_2$ compounds.

Figure 3(a) represents the magnetization (*M*) versus applied field (*H*) plots for the pure $MgB_2$ compound being used in the present study. The *M*(*H*) plots are taken in applied fields of $-7\ T \leq H \leq\ 7T$ and at $T = 5$, 10, 20, 25, 30 and 35 K. The units of magnetization (*M*) used in figure 3(a) are emu/cm$^3$, which are calculated as the usual process being given in detail in ref. 11 for calculating the $J_c$ by Bean model. The *M*(*H*) plots width is more or less same at $T = 5$, 10 K, and later decrease with increase in temperatures to 20, 25, 30 and 35 K. The *M*(*H*) loops are practically open till the applied fields ($H_m^*$) of up to say 4.5 Tesla for $T = 5$, 10 K and at progressively lower fields for higher temperatures. Before we calculate the $J_c$ from these *M*(*H*) loops for pure $MgB_2$ of present study, let us discuss the shapes and size of the various nano-$Co_3O_4$ doped $MgB_2$ compounds.

The *M*(*H*) plots being taken in applied fields of $-7\ T \leq H \leq\ 7T$ and at $T = 5$, 10, 20, 25 and 30 K for 2% nano-$Co_3O_4$ doped $MgB_2$ are given in figure 3 (b). The units of magnetization (*M*) used in figure 3(b) are also in emu/cm$^3$ to facilitate the comparison with figure 3(a) and to be used for $J_c$ calculation by Bean model. There are few peculiar things to be noted in regards to *M*(*H*) plots in figure 3 (b); first the absolute value of *M*(emu/cm$^3$) is nearly 1.5 times when compared to *M*(*H*) plots in figure 3 (a) at mostly all fields and temperatures, secondly strong fluxiod jumps are seen below applied fields of ± 1 T at $T = 5$, 10 K. For sack of brevity one can say that the magnetizations seems to be nearly invariant below the applied fields of ± 1 T at $T$ = 5, 10 K. Worth noting also is another interesting fact, that $H_m^*$ being defined in earlier paragraph is close to 7 T at 5 K for the 2% nano-$Co_3O_4$ doped $MgB_2$. It is concluded from the comparison of figure 3(a) and 3(b) that the magnetization response of 2% nano-$Co_3O_4$ doped $MgB_2$ is far superior to the pristine $MgB_2$.

One questions need to be answered here is about the true nature of fluxiod-jumps in the $M(H)$ plots of 2% nano-$Co_3O_4$ doped $MgB_2$. We found in various earlier reports, that when the $J_c$ of $MgB_2$ is high enough to the tune of $10^7$ A/cm$^2$ or above, such fluxiod jumps are seen irrespective of the type of pinning centers used [7, 11, 19]. Ironically not all reports comment on this aspect, but what was clear from their fluctuating $J_c(H)$ plots below say 1 T, that they really had observed fluxiod jumps in respective $M(H)$ plots. For this region only, we intended to show the $M(H)$ plots in figures 3(a) and 3(b). The fluxiod jumps (bit less pronounced) similar to that in figure 3(b) are also observed in $M(H)$ plots of 4% nano-$Co_3O_4$ doped $MgB_2$ sample, plots not shown. Interestingly enough the shape of $M(H)$ plots of 6% nano-$Co_3O_4$ doped $MgB_2$ sample are very similar to that as for pure $MgB_2$ in figure 3(a). Also it will be noted soon in next section, that $J_c$ is highest for the 2% nano-$Co_3O_4$ doped $MgB_2$ in comparison to all other samples and is least for 6% nano-$Co_3O_4$ doped sample. Importantly the fluxiod jumps are also much pronounced in the 2% nano-$Co_3O_4$ doped sample having highest $J_c$ at all fields and studied temperatures and not seen at all in pristine or 6% doped sample.

The critical current density ($J_c$) being determined by Beans model is given as following:

$$J_c = (30 \text{ x } \Delta M/d) \text{ x } (10/4\pi) \text{ (A/cm}^2) \qquad (1)$$

Here, $\Delta M$ is the width of the $M(H)$ loop at a given field and temperature and $d$ is the average grain size of the material. In fact formula used in Eqn.1 is for cylindrical sample being used in the magnetization and $d$ is the diameter of the same. Though we managed to use the specimen in magnetization measurements in nearly cylindrical shape, the very porous nature of the samples, warrant the use of average grain size as $d$. This is a usual practice in case of porous polycrystalline $MgB_2$ compounds [4-11]. In any case, we had discussed this issue in detail in our recent earlier communication on Mg/MgO added $MgB_2$ compounds [11]. The $J_c$, thus calculated with the above reasoning, is shown in figure 4 for all studied samples. The fluxiod jumps region, which was found for 2% and 4% nano-$Co_3O_4$ doped samples, is marked on the plots. It is seen that the $J_c$ (~$10^8$ A/cm$^2$, at $T \leq 10$ K, $H \leq 1$ T) of 2% nano-$Co_3O_4$ doped sample is nearly an order of magnitude higher than as for pristine $MgB_2$ at all fields and temperatures. Similarly the $J_c$ of 4% nano-$Co_3O_4$ doped sample is also higher than as for pristine $MgB_2$, but is less than that as for 2% nano-$Co_3O_4$ doped sample. Finally we see that the $J_c$ of 6% nano-$Co_3O_4$ doped sample is least in all samples including pristine $MgB_2$ at nearly all fields and temperatures.

An interesting question need to be answered is now the possible reason behind the interesting observations of *PME* and fluxiod jumps being accompanied with high $J_c$ of the nano-$Co_3O_4$ doped $MgB_2$ compounds. For this we show the X-ray diffraction (XRD) patterns of the various nano-$Co_3O_4$ doped $MgB_2$ compounds in figure 5. It is evident from figure 5 that all the samples are crystallized in required hexagonal Bravais lattice with lattice parameters of $a = 3.08$ Å, and $c = 3.52$ Å. For pristine sample small quantities of Mg and MgO are also seen, which are comparatively very small to that as for reported Mg turnings added $MgB_2$ [11]. In fact the pristine $MgB_2$ is present study is nearly pure phase. Perhaps this is the reason, that present $MgB_2$ does not have higher critical current density coupled with fluxiod jumps/constant $J_c$ regions, like as in ref. 11. Phase pure $MgB_2$, which are devoid of added pinning centers do generally possess a low critical current. Nearly unchanged lattice parameters of the nano-$Co_3O_4$ doped $MgB_2$ compounds are an indication to the fact that Co has not entered the lattice to result in $Mg_{1-x}Co_xB_2$. In fact as we go on increasing the nano-$Co_3O_4$ content an extra XRD peak starts appearing close to main intensity peak due to the presence of un-reacted $Co_2O_3$ in the compound.

We feel the abundance of un-reacted $Co_2O_3$ in nano-$Co_3O_4$ doped $MgB_2$ compounds acts as effective pinning centers and hence resulting in an increased $J_c$ coupled with fluxiod jumps. Also this could give rise to the *PME* in terms of the pi junctions of $MgB_2/Co_2O_3/MgB_2$. There also seems to be an optimum of the doped nano- $Co_3O_4$ in the pristine $MgB_2$ to give highest $J_c$, which is around 2%-4%. For higher percentage of nano- $Co_3O_4$ ($\geq 6\%$) the $J_c$ goes down and the fluxiod jumps are not seen. The distribution of pinning centers in a material is equally important as their density, to let them act as effective pinning centers. Micro-structure studies of such materials in important. Though, we feel that the mapping of the distribution of pinning centers in a bulk material could not be very informative. We believe the aligned thin films of nano-particle doped $MgB_2$ are warranted to study the effective distribution of pinning centers. The fluxiod jumps are seen only for higher $J_c$ samples and are commented on earlier in some reports [7, 11]. The fluxiod jumps are suppose to occur only in case of extremely high critical currents and very low heat capacity, resulting in localized motion of magnetic flux [7, 20]. Further these jumps could be nearly stabilized if the fields intervals applied are small [11]. The importance of intrinsic fluxiod jumps and high $J_c$ are discussed earlier in detail in ref. 7 and 11. A more precise recent detailed discussion on the appearence of fluxiod jumps in terms of magnetic field dependent contribution of specific heat of $MgB_2$ in superconducting state is highlighted in ref. 21. The unusual structure of the flux jumps was found, which could be explained on the basis of specific heat and magnetic properties of this class of compounds [21].

## CONCLUSIONS and SUMMARY

In conclusion, we have been able to synthesize nano-$Co_3O_4$ doped $MgB_2$ compounds by vacuum annealing process. The critical density of the prestine compound is found to increase with nano-$Co_3O_4$ doping, with an optimum $J_c$ at 2%-4% doping level. Further the higher $J_c$ (~$10^8$ A/cm$^2$, at $T \leq 10$ K, $H \leq 1$ T) is found to be coupled with the fluxiod jumps in low field regions. The abundance of un-reacted $Co_2O_3$ in nano-$Co_3O_4$ doped $MgB_2$ compounds must be acting as effective pinning centers. Also the presence of $Co_2O_3$ could give rise to the *PME* in terms of the pi junctions of $MgB_2/Co_2O_3/MgB_2$.

## ACKNOWLEDGEMENT

Authors from NPL appreciate the interest and advice of Prof. Vikram Kumar (Director) NPL in the present work. The work is partly supported by INSA-JSPS bilateral exchange program.

**FIGURE CAPTIONS**

Fig. 1 Resistiviy ($\rho$) versus temperature (*T*) plots for 0, 2, 4 and 6% nano-$Co_3O_4$ doped $MgB_2$ system.

Fig. 2 Dc magnetic susceptibility versus temperature $\chi(T)$ plots in both Zero-field-cooled (*ZFC*) and Field-cooled (*FC*) situations for 0, 2, 4 and 6% nano-$Co_3O_4$ doped $MgB_2$ system.

Fig. (3a) Magnetic hysteresis *M*(*H*) loops plots for 750 $^0$C annealed $MgB_2$ pristine compound at 5, 10, 20, 25, 30 and 35 K with applied fields (*H*) of up to ± 7 Tesla.

Fig. (3b) Magnetic hysteresis *M*(*H*) loops plots for 750 $^0$C annealed 2% nano-$Co_3O_4$ doped $MgB_2$ at *T* = 5, 10, 20, 25, 30 and 35 K with applied fields (*H*) of up to ± 7 Tesla.

Fig. (4) Critical current density $J_c$ (*H*) plots at various temperatures and applied fields of up to 7 Tesla for 0, 2, 4 and 6% nano-$Co_3O_4$ doped $MgB_2$ system.

Fig. 5 X-ray diffraction pattern for 750 $^0$C annealed and nano-$Co_3O_4$ doped $MgB_2$ compounds.

Figure 1

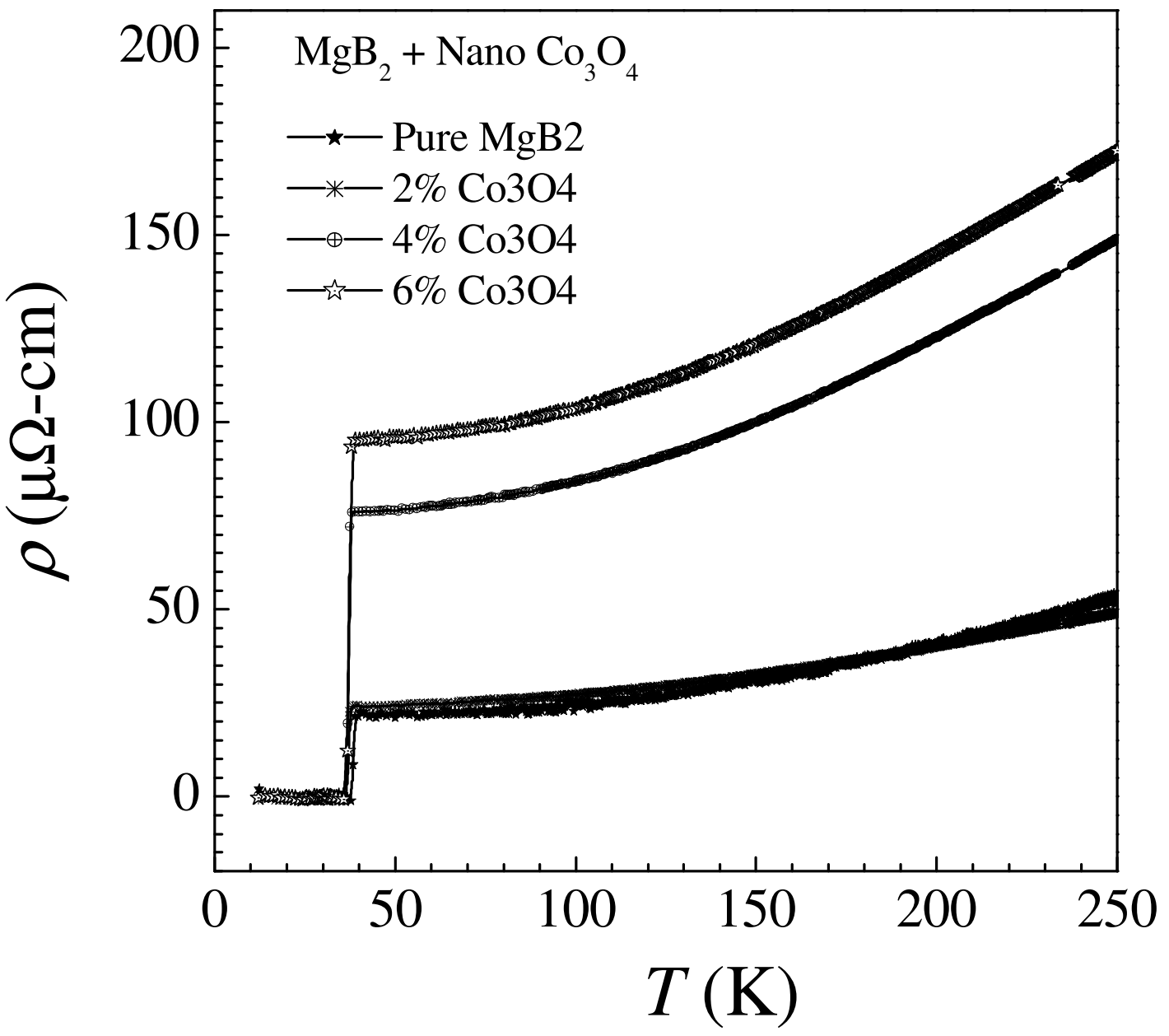

Figure 2

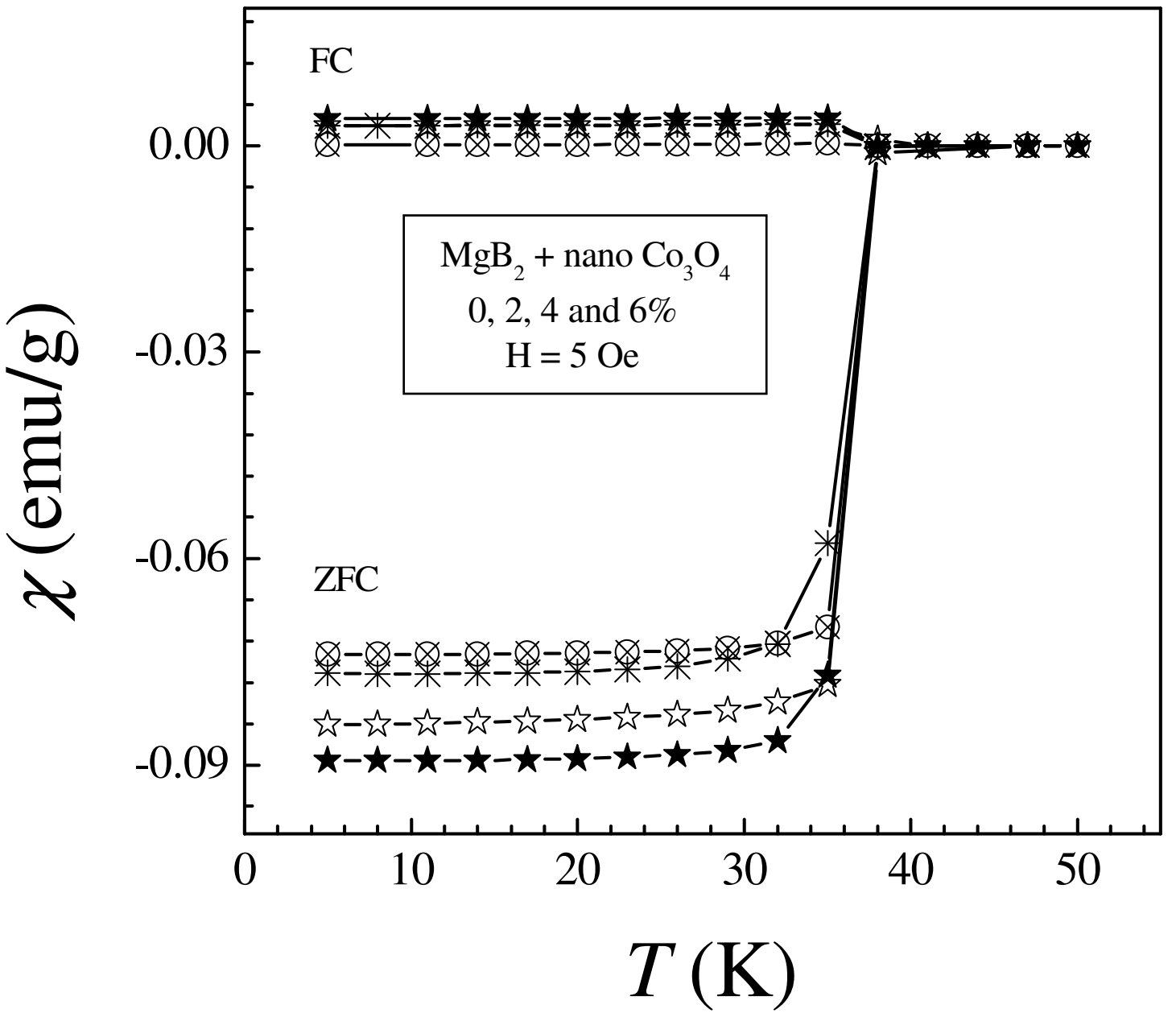

Figure 3 (a)

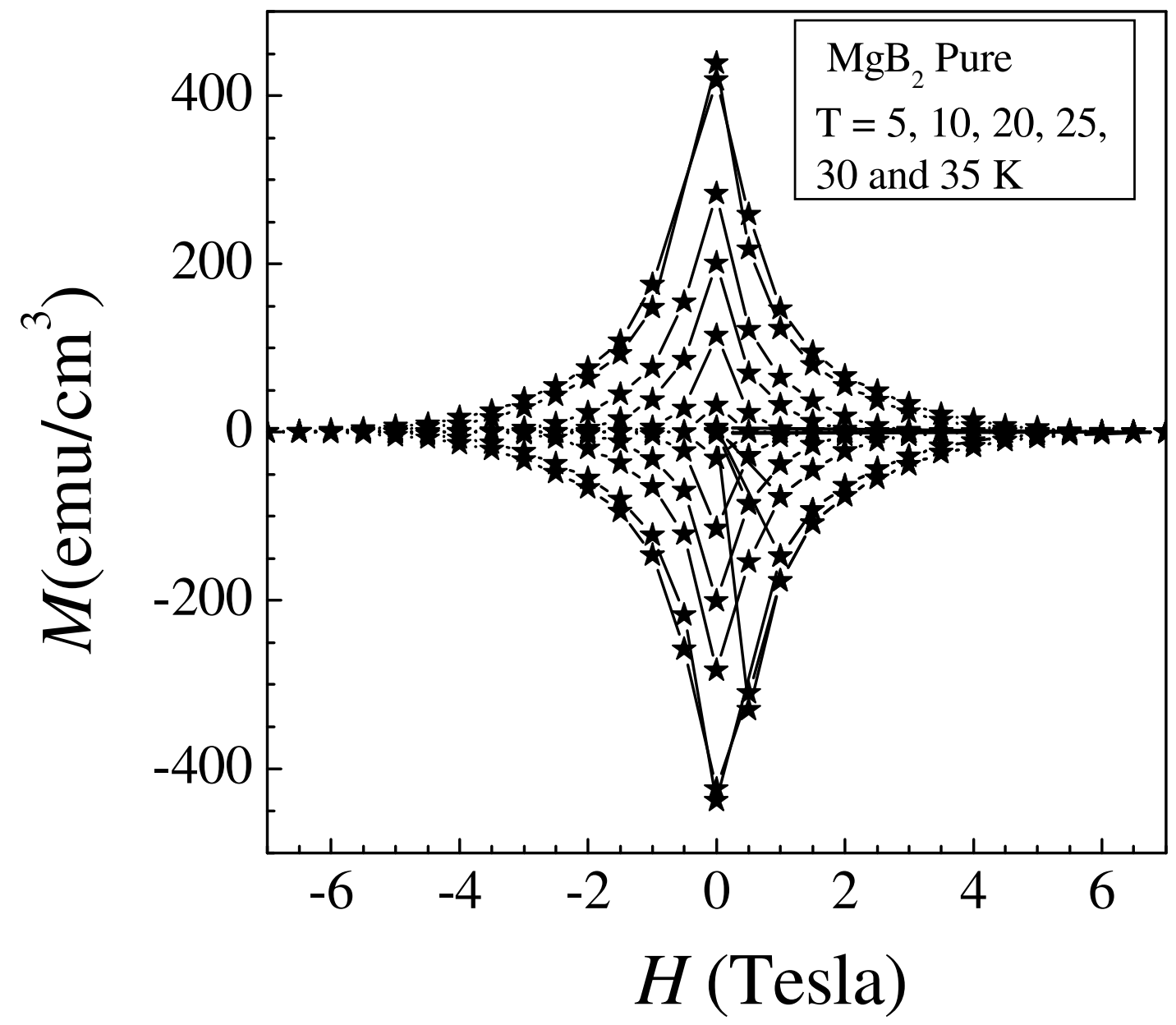

Figure 3(b)

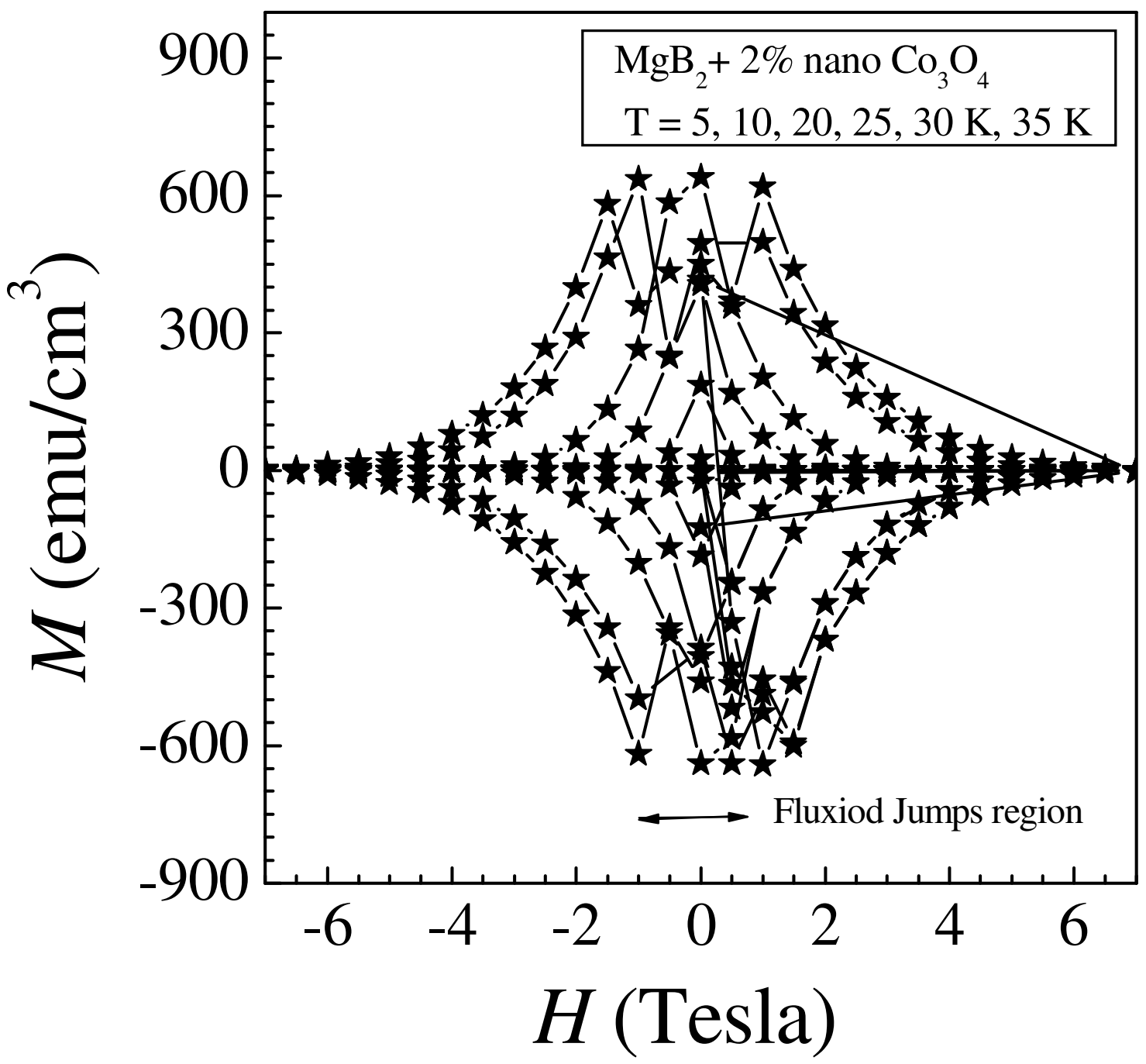

Figure 4

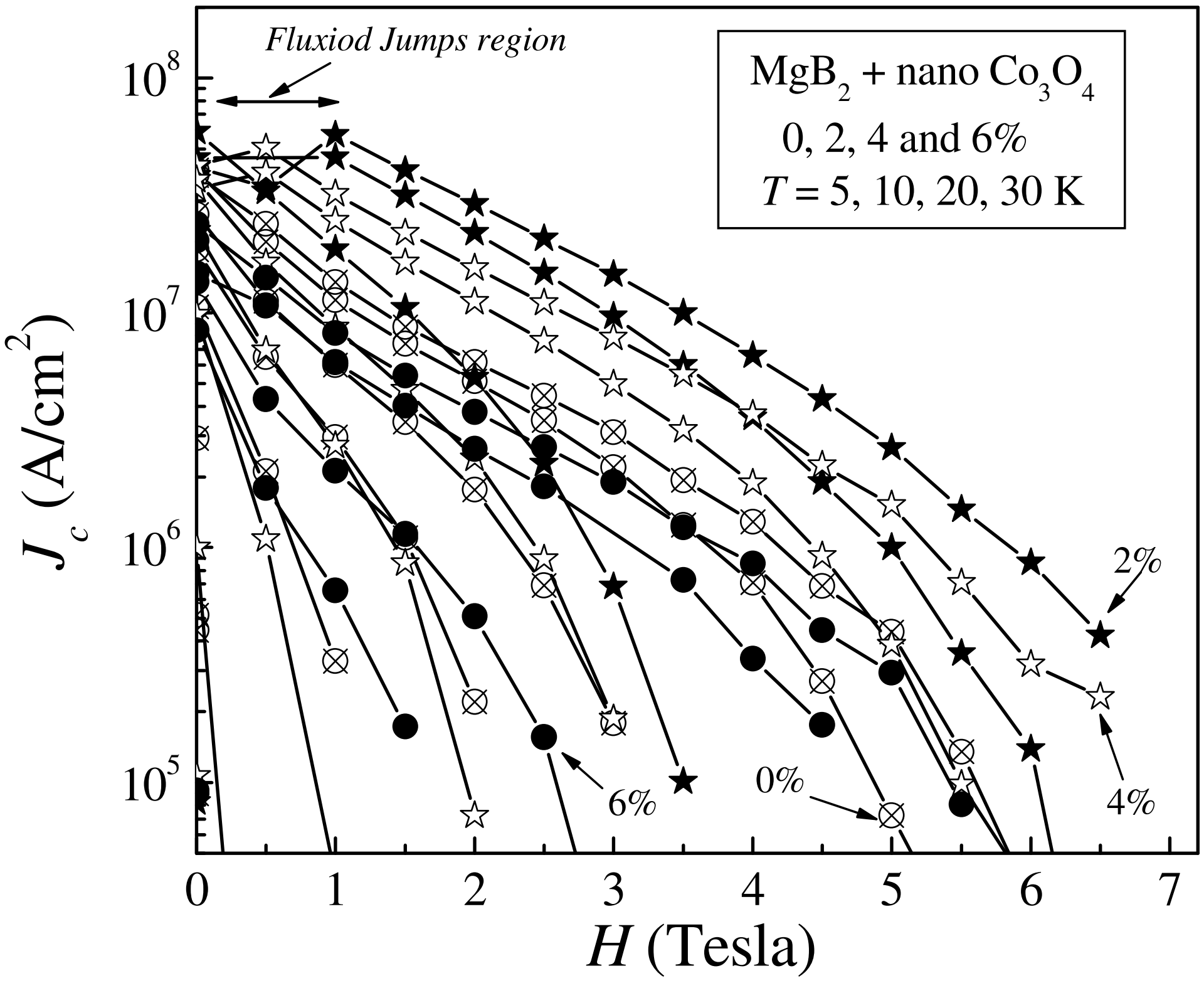

Figure 5

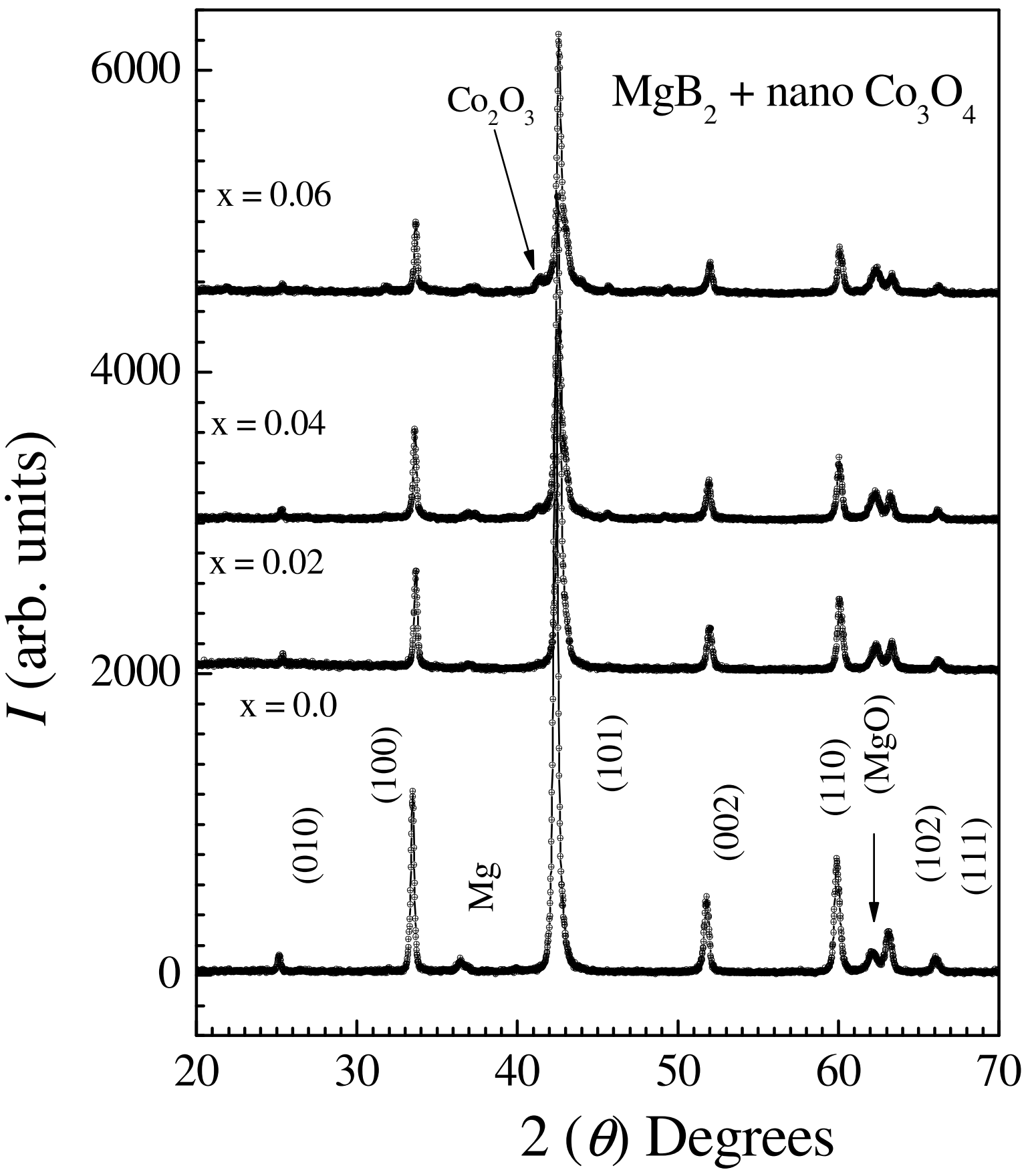